\documentclass[a4paper]{jpconf}
\usepackage{graphicx}
\begin{document}
\title{\boldmath Physics at Super$B$}

\author{Tim Gershon}
\address{Department of Physics, University of Warwick, Coventry, CV4 7AL, UK}
\ead{T.J.Gershon@warwick.ac.uk}

\begin{abstract}
  Flavour will play a crucial role in understanding 
  physics beyond the Standard Model.
  Progress in developing a future programme to investigate 
  this central area of particle physics has recently passed a milestone,
  with the completion of the conceptual design report for Super$B$,
  a very high luminosity, asymmetric $e^+e^-$ collider.
  This article summarizes the important role of Super$B$
  in understanding new physics in the LHC era.
\end{abstract}

The major challenge facing particle physics in the next decade is to 
go beyond the Standard Model (SM) of elementary particles.
Historically, progress in the field has been achieved through advances on two 
parallel approaches -- the energy frontier and the luminosity frontier.
While the LHC will soon deliver the much anticipated leap forwards
in available centre-of-mass energy,
a comparable advance from the current world record luminosities
achieved by the $B$ factories will allow complementary research 
into new physics (NP).\footnote{
  The development of high luminosity machines is also clearly 
  beneficial for the health of accelerator-based physics,
  as discussed in the EPS-ECFA joint session at EPS2007.
}
This is a simple consequence of the fact that quantum physics 
allows for the virtual production of heavy particles,
which influence processes at energies much below their masses.
Therefore, to investigate NP through quantum effects,
high precision, rather than high energy, is required.

The flavour sector is well-suited to search for quantum effects of NP
since flavour changing neutral currents, 
neutral meson-antimeson mixing and $CP$ violation 
all occur at the loop level in the SM,
with quark flavour violation further suppressed by the small mixing angles,
Since there is no {\it a priori} reason for NP to share these features, 
the quark sector is potentially subject to large NP effects.
Similar arguments apply in the lepton sector,
where the theoretical interest is even greater,
since the physics behind neutrino oscillations remains an open question.

Many observables in the flavour sector 
are generically sensitive to NP effects,
including rates and asymmetries of 
rare leptonic or loop-induced $K$, $D$ and $B$ decays,
mixing and $CP$ violating phenomena in the 
$K^0$, $D^0$, $B_d^0$ and $B_s^0$ systems,
electric and magnetic dipole moments of charged leptons 
and lepton flavour violating $\mu$ and $\tau$ decays.
While certain models focus attention onto particular channels,
there is no single ``golden mode'' --
rather, the flavour sector can be thought of as a treasure chest
of NP-sensitive observables.
Indeed, the plethora of measurements that can be made adds significantly
to the physics programme, enhancing the sensitivity to NP.
Moreover, correlations between observables can distinguish between
different NP models.

A coherent programme for particle physics research in the next decade
should therefore allow as many flavour observables as possible to be studied.
No single experimental facility can cover them all.
However, a ``Super Flavour Factory'',
{\it i.e.} a high luminosity, asymmetric $e^+e^-$ collider
has a very wide-reaching potential, allowing for comprehensive studies 
of charm ($D^0$, $D^+$ and $D_s^+$) and beauty ($B^+$ and $B_d^0$) mesons 
as well as tau leptons.
This facility is thus established as the foremost priority for flavour physics,
yet it must be emphasized that it is complementary
to dedicated experiments in the kaon and muon sectors,
while observables related to $B_s^0$ meson oscillations 
can be better measured elsewhere.

There is insufficient space in this article 
to describe even a fraction of the most important measurements
from the broad physics programme of Super$B$~\cite{Bona:2007qt}.
The machine design has several apparent advantages compared to 
conventional upgrades of the existing $B$ factories~\cite{Hashimoto:2004sm}.
These include:
peak instantaneous luminosity in excess of $10^{36} \ {\rm cm}^2 {\rm s}^{-1}$,
providing for nominal integrated luminosity of $75 \ {\rm ab}^{-1}$ after 
five years of operation;
detector backgrounds and power consumption 
comparable to the current $B$ factories;
flexible centre-of-mass energy
and an option for beam polarization
that extend the reach for several interesting observables.
These points notwithstanding, 
for the physics at the $\Upsilon(4{\rm S})$ there is much in common 
between the programmes of machines with different designs~\cite{Akeroyd:2004mj}.

\begin{table}[tb]
  \caption{
    Expected precision of some of the most important measurements
    that can be performed at Super$B$ with $75 \ {\rm ab}^{-1}$.
    Numbers quoted as percentages are relative precisions.
    Values given for rare tau decays are the 
    $90\%$ confidence level upper limits expected in the absence of signal.
    Measurements marked ($\dagger$) will be systematics limited;
    those marked ($\ast$) will be theoretically limited.
    In many of these cases, 
    there exist data driven methods of reducing the errors.
    \smallskip
  }
  \label{tab:precision}

  \begin{tabular}{cc}
    \begin{minipage}{0.49\textwidth}

      \begin{center}
        \begin{tabular}{lc}
          \hline
          Observable               & Precision \\
          \hline
          $\sin(2\beta)$ ($J/\psi\,K^0$)    &     0.005 ($\dagger$) \\
          $\alpha$ ($\pi\pi, \rho\pi, \rho\rho$ combined) &  $1$--$2^\circ$ ($\ast$) \\
          $\gamma$ ($B \to DK$, combined)   &     $1$--$2^\circ$  \\
          $\left| V_{ub} \right|$ (inclusive) &       $2.0\%$ ($\ast$)   \\
          \hline
          $S(\phi K^0)$                     &        0.02 ($\ast$)     \\
          $S(\eta^\prime K^0)$               &        0.01 ($\ast$)     \\
          $S(K_S^0K_S^0K_S^0)$               &        0.02 ($\ast$)     \\
          \hline
          $\phi_D$                          &      $1$--$3^\circ$      \\
          \hline
          ${\cal B}(\tau \to \mu\,\gamma)$        &  $2 \times 10^{-9}$     \\
          ${\cal B}(\tau \to \mu\, \mu\, \mu)$    &  $2 \times 10^{-10}$  \\
          \hline
        \end{tabular}
      \end{center}
    \end{minipage}

    &

    \begin{minipage}{0.49\textwidth}

      \begin{center}
        \begin{tabular}{lc}
          \hline
          Observable               & Precision \\
          \hline
          ${\cal B}(B \to \tau \nu)$        &       $ 4\%$ ($\dagger$) \\
          ${\cal B}(B \to \mu \nu)$         &       $5\%$             \\
          ${\cal B}(B \to D \tau \nu)$      &       $ 2\%$             \\
          \hline
          ${\cal B}(B \to \rho \gamma$)     &       $  3\%$ ($\dagger$) \\
          $A_{CP}(b \to s \gamma)$           &       $0.004$ ($\dagger$) \\
          $A_{CP}(b \to (s+d) \gamma)$       &       $0.006$ ($\dagger$)  \\
          $S(K_S^0\pi^0\gamma)$             &       $0.02$ ($\ast$)     \\
          $S(\rho^0\gamma)$                 &       $0.10$              \\
          $A^{FB}(B \to X_s \ell \ell) s_0$  &         $5\%$             \\
          ${\cal B}(B \to K \nu \overline{\nu})$   &         $20\%$            \\
          \hline
        \end{tabular}
      \end{center}

    \end{minipage}
  \end{tabular}
\end{table}

Tab.~\ref{tab:precision} summarizes the sensitivity of Super$B$
for some of the most important observables.
The differences compared to the current $B$ factories are striking.
The SM CKM parameters can be precisely measured,
even in the presence of arbitrary NP contributions 
(assuming progress in the precision of 
lattice QCD calculations~\cite{Bona:2007qt}),
providing the necessary reference point for interpretation of NP signals.
At Super$B$, measurements of known rare processes such as 
$b\to s \gamma$ or $CP$ violation in hadronic $b\to s$ penguin transitions
such as $B^0 \to \phi K_S^0$ will be advanced to unprecedented precision.
Channels which are just being observed in the existing data,
such as $B^0 \to \rho^0\gamma$, $B^+ \to \tau^+\nu_\tau$ 
and $B \to D^{(*)}\tau\nu$ will become precision measurements.
Furthermore, detailed studies of decay distributions and asymmetries
that cannot be performed with the present statistics
will significantly improve the sensitivity to NP.
A salient example lies in $D^0$--$\bar{D}^0$ oscillations:
the current evidence for charm mixing, 
which cannot be interpreted in terms of NP, 
opens the door for precise measurements of 
the $CP$ violating phase in charm mixing ($\phi_D$), which is known to 
be zero in the Standard Model with negligible uncertainty.
In addition, these measurements will be accompanied by 
dramatic discoveries of new modes,
including decays such as 
$B^+ \to K^+ \nu \bar{\nu}$ and $B^+ \to \pi^+ \ell^+ \ell^-$, 
which are the signatures of the theoretically clean quark level processes
$b \to s \nu \bar{\nu}$ and $b\to d \ell^+\ell^-$ respectively.

\begin{figure}[tb]
  \begin{picture}(500,190) 
    \put(20,180){
      \includegraphics[angle=-90, bb=80 150 520 620, width=0.42\textwidth]{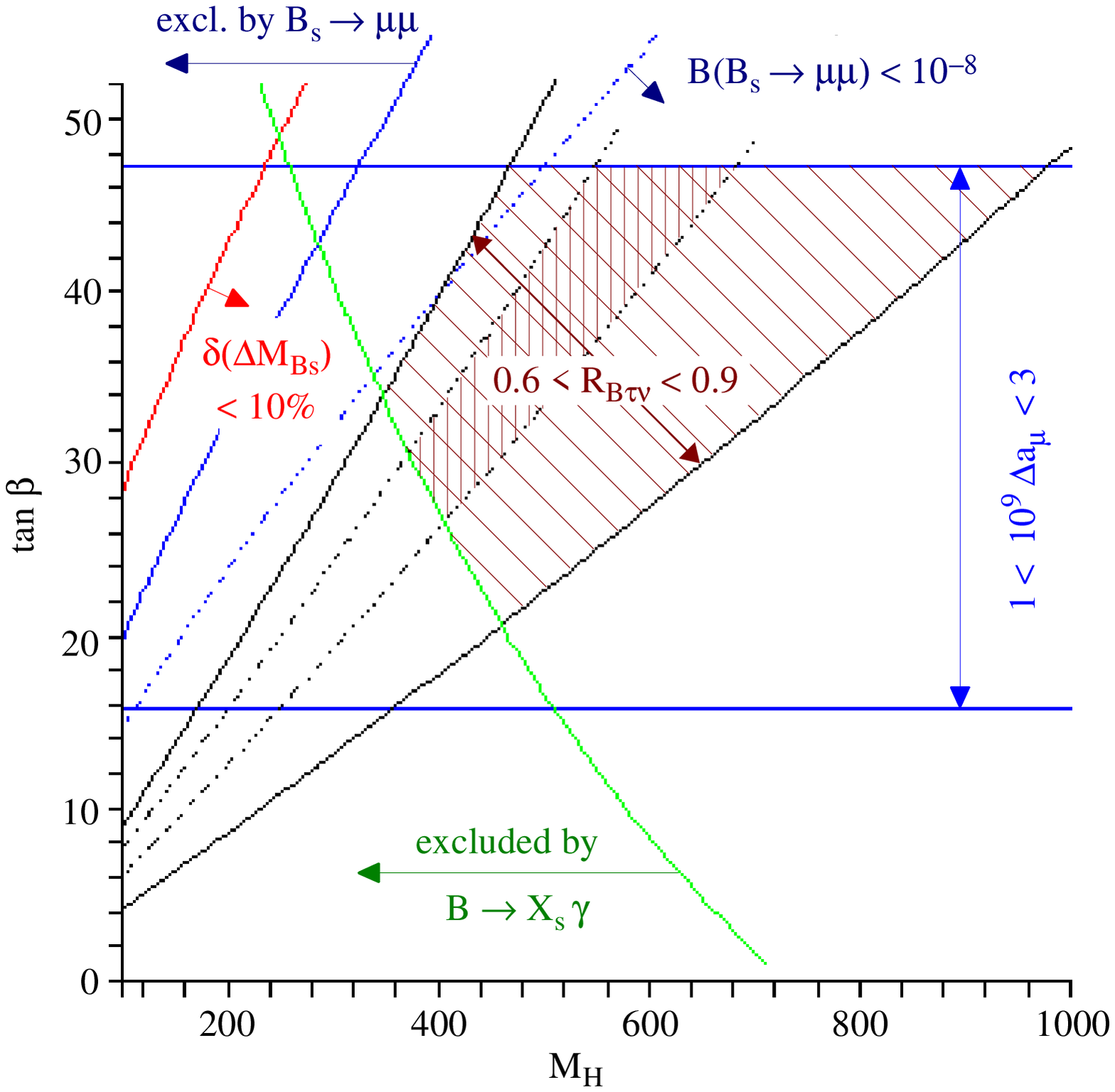}
    }
    \put(170,30){(a)}
    \put(230,-10){
      \includegraphics[width=0.42\textwidth]{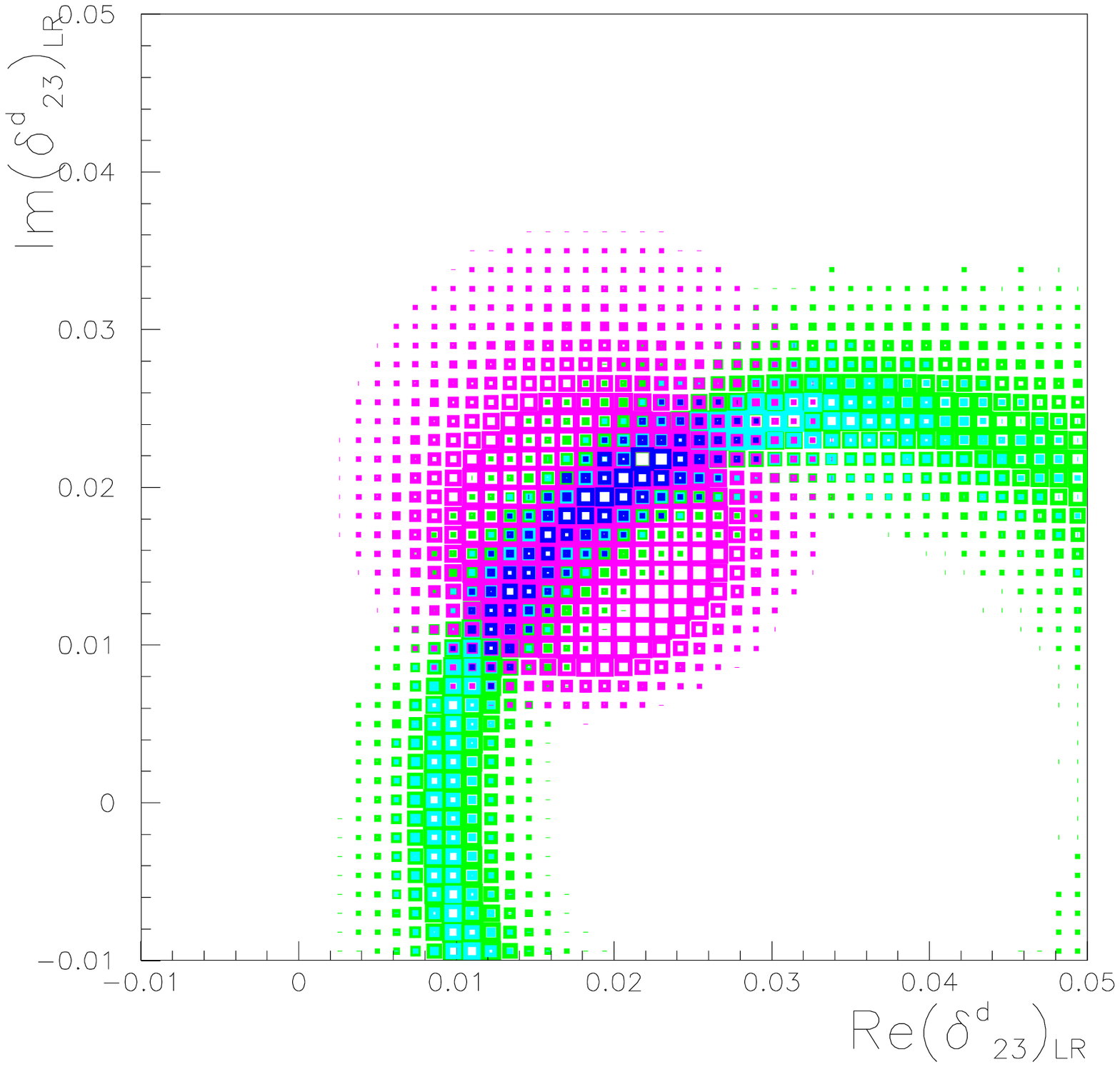}
    }
    \put(380,30){(b)}
  \end{picture}
  \caption{
    (a) Constraints in the $\tan\beta$--$M_{H^+}$ plane
    obtained from 
    $\Delta m_s$, ${\cal B}(B_s^0 \to \mu^+\mu^-)$, $\Delta a_\mu$,
    ${\cal B}(B \to X_s \gamma)$ and ${\cal B}(B^+ \to \tau^+ \nu_\tau)$~\cite{Isidori:2006pk}.
    (b) Constraints on the mass insertion parameter $(\delta_{23}^d)_{LR}$
    that could be obtained by Super$B$, using measurements of 
    ${\cal B}(B \to X_s \gamma)$ (green), ${\cal B}(B \to X_s l^+l^-)$ (cyan),
    $A_{CP}(B \to X_s \gamma)$ (magenta) 
    and all combined (blue)~\cite{Bona:2007qt}.
  }
  \label{isidori}
\end{figure}

There has been much recent activity exploring the interplay
between flavour observables and measurements 
that will be made at the LHC~\cite{flavourLHC}.
This work constitutes preliminary steps towards 
the ultimate challenge of reconstructing the new physics Lagrangian 
from the combination of experimental inputs.
However, one outcome is particularly relevant for today.

Any new-physics model, established at the TeV scale 
to solve the gauge hierarchy problem, includes new flavoured particles
which may be discovered at the LHC.
Since these must couple to the SM particles, the NP cannot be flavour blind,
and flavour observables must be affected.
Indeed, this is the origin of the ``flavour problem'' --
{\it i.e.} the nonappearance to date of NP in flavour observables
such as neutral meson mixing parameters.
A solution to this problem can be found in 
the concept of minimal flavour violation (MFV),
whereby NP follows the SM pattern of flavour- and $CP$- violation 
encoded in the CKM quark mixing matrix~\cite{D'Ambrosio:2002ex}.

Although the MFV scenario is far from established
(and could only be verified by precise measurements of flavour observables),
it is nonetheless very useful to define the minimum NP-sensitivity of flavour.
It has been shown that even in such an unfavourable case, 
Super$B$ can still detect deviations from the SM 
for NP particle masses up to at least $600 \ {\rm GeV}$~\cite{Bona:2007qt}.
At large values of $\tan\beta$ (the ratio of Higgs vacuum expectation values),
this bound is raised by at least a factor of three.
A good example of this is shown in Fig.~\ref{isidori}(a),
where the constraints from various flavour observables 
in the $\tan\beta$--$M_{H^+}$ plane are shown in a MFV scenario,
which could be realised in a two Higgs doublet model or in the MSSM~\cite{Hou:1992sy,Isidori:2006pk}.
Once other possible extensions to MFV are considered,
the phenomenology accessed by Super$B$ rapidly becomes much richer,
allowing NP flavour couplings that are not accessible at LHC to be measured.
An example is shown in Fig.~\ref{isidori}(b),
where one of the MSSM mass insertion parameters
can be completely determined from Super$B$ measurements~\cite{Bona:2007qt}.
Similar conclusions hold for all NP parameters related to 
$b \to s$ or $b \to d$ transitions.

On the other hand, it is entirely possible that the masses of new particles
may be out of reach of the LHC.
In this case, Super$B$ can be used to probe much higher mass scales,
since NP models with unsuppressed flavour couplings 
can cause visible effects even if the new particles have masses of
${\cal O}(100 \ {\rm TeV})$ or more.
Thus, Super$B$ can provide essential complementary information on NP,
regardless of whether or not it is discovered by LHC.

\begin{figure}[tb]
  \centering
  \includegraphics[angle=-90, width=0.55\textwidth]{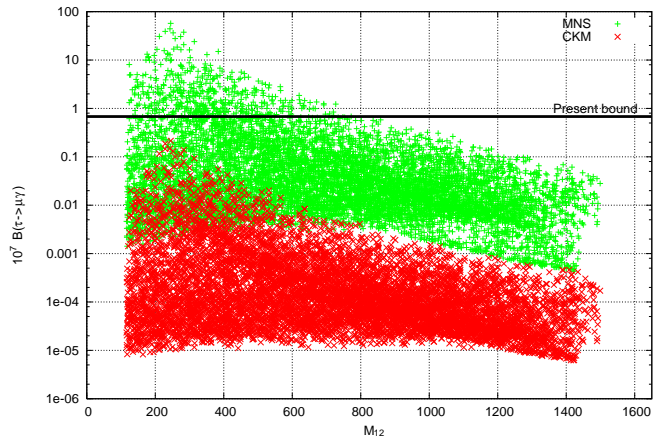}
  \caption{
    ${\cal B}(\tau\to \mu\gamma)$ in units of $10^{-7}$ {\it vs.}
    the high energy universal gaugino mass ($M_{1/2}$)
    within a $SO(10)$ framework~\cite{Calibbi:2006nq}.
    Green and red points correspond to the scenarios where 
    $Y_{\nu} = U_{\rm PMNS}$ and where $Y_{\nu} = V_{\rm CKM}$, respectively.
    The thick horizontal line denotes the present experimental sensitivity,
    which would be improved by almost two orders of magnitude at Super$B$.
  }
  \label{masieroLFV06}
\end{figure}

It is important to emphasise that Super$B$ is not limited 
to the study of $B$ mesons, but is really a Super Flavour Factory.
In particular, searches for lepton flavour violating (LFV) tau decays
provide an interesting link with neutrino oscillations.
In models with Majorana neutrinos, 
LFV processes occur at rates that depend on the heavy masses.
Moreover, the pattern of neutrino mixing angles suggests that
$\tau \to \mu \gamma$ decay should have the largest LFV rate,
while that for $\mu \to e \gamma$ may depend on 
the as-yet-unmeasured value of $\theta_{13}$~\cite{Antusch:2006vw}.
Furthermore, one of the most interesting possibilities in NP models is 
the unification of quark and lepton sectors.
As an example, Fig.~\ref{masieroLFV06} shows 
the prediction for ${\cal B}(\tau\to\mu\gamma)$ 
within a supersymmetric SO(10) framework~\cite{Calibbi:2006nq}.
as a function of the gaugino mass $M_{1/2}$.
The scenarios where the source of LFV violation is governed by 
neutrino mass matrix $Y_{\nu}=U_{\rm PMNS}$ and where $Y_{\nu} = V_{\rm CKM}$ 
can be distinguished.
Super$B$ has sensitivity for tau physics that is superior to any 
other existing or proposed experiment, 
even in channels that appear well-suited to the LHC, 
such as $\tau \to \mu\mu\mu$.

To summarize, the case for flavour physics in the LHC era is compelling.
A Super Flavour Factory is the ideal tool 
to cover as much as possible of the wide range of interesting physics 
in both quark and lepton sectors.
Many more details on the physics case can be found in 
the conceptual design report for Super$B$~\cite{Bona:2007qt}.

\ack
I would like to thank the conveners of the 
Flavour Physics and $CP$ Violation parallel session
and the organizers of the EPS2007 conference for the invitation 
to present this work.
I would also like to thank
T.~Browder, M.~Ciuchini, T.~Hurth, M.~Hazumi and A.~Stocchi.

\end{document}